\documentclass[pre,floatfix,twocolumn]{revtex4}  

\usepackage{epsfig}
\usepackage{graphicx}

\begin{document}

\author{Brian Utter}
\email{utter@phy.duke.edu}
\author{E. Bodenschatz}
\affiliation{Laboratory of Atomic and Solid State Physics, Cornell University,
	Ithaca, NY  14853}
\date{\today}
\title{Doublon Growth in Solidification}

\renewcommand{\textfraction}{0.15}
\renewcommand{\topfraction}{0.95}
\renewcommand{\bottomfraction}{0.95}
\setcounter{bottomnumber}{2} 
\setcounter{topnumber}{2}
\renewcommand{\floatpagefraction}{0.75}

\newcommand{\ea}{{\it et al.}}
\newcommand{\lbf}{\large \bf}

\begin{abstract}		
We present experiments on the doublon growth morphology in directional 
solidification.  Samples used are succinonitrile 
with small amounts of
poly(ethylene oxide), acetone, or camphor as the solute.  
Doublons, or symmetry-broken dendrites, 
are generic diffusion-limited growth structures 
expected at large undercooling and low anisotropy.
Low anisotropy growth is achieved by selecting a grain near
the $\{$111$\}$ plane leading to either seaweed (dense branching morphology)
or doublon growth depending 
on experimental parameters.
We find selection of doublons to be strongly 
dependent on solute 
concentration and sample orientation. 
Doublons are selected 
at low concentrations (low solutal undercooling) 
in contrast to the prediction of doublons at large thermal undercooling in 
pure materials. 
Doublons also exhibit preferred growth 
directions and changing the orientation of a specific doublonic 
grain changes the character and stability of the doublons.  
We observe transitions between seaweed and doublon growth 
with changes in concentration and sample orientation.  
\end{abstract}


\maketitle


\section{Introduction}

Over the past 20 years, surface tension anisotropy has been discovered to be 
fundamental in determining solidification morphology.  
Perhaps the greatest success was the discovery that 
anisotropy is required for the formation of stable cells and dendrites
through a microscopic solvability condition \cite{reviews-PRE-01}.
In contrast, an isotropic surface tension leads to complicated, 
tip splitting growth known as 
dense branching morphology \cite{Ben-Jacob.ea:86:Formation} or seaweed 
growth \cite{ihle-mk-93-4}. 
This is a generic feature of diffusion limited growth 
and exists in a variety of systems
under isotropic or weakly anisotropic 
conditions, such as viscous fingering, bacterial colony growth, and 
electrodeposition  \cite{low-anis-refs}.  

\begin{figure}
\includegraphics[width=3.3in]{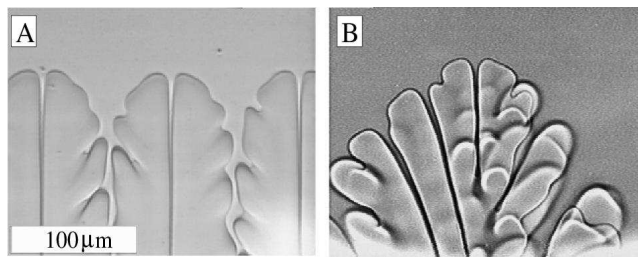}
\caption{
(A) Stable doublons in 0.5\% PEO-SCN at a growth rate 
V = 22 $\mu$m/s. (B) Seaweed growth
in 0.25\% PEO-SCN at the same groeth rate.  A transient 
doublon develops before breaking apart.
}
\label{doublon-seaweed}
\end{figure}

It was predicted for the solidification of a pure material  
that a transition from fractal to compact seaweed occurs
with increasing undercooling \cite{brener-96-8}.
For a pure material, thermal undercooling 
increases with growth speed for both dendrites and
doublons.
At low speeds, perturbations 
at the tip lead to tip splitting while at higher speeds they  
convect away
\cite{brener-96-8}.   The basis of the compact seaweed is the doublon, 
or symmetry-broken dendrite,
in which two asymmetric cells grow cooperatively such that 
there is a parabolic envelope over the pair of cells and a thin liquid 
gap of well-defined size 
separating them \cite{Akamatsu.ea:95:Symmetry-broken}. 
An example of this is shown in Fig.~\ref{doublon-seaweed}A.
It is somewhat counterintuitive to observe 
random seaweed patterns at low driving force and oriented doublon growth 
at large driving force.
The isotropy in surface tension 
is revealed in that although doublons have a clear 
orientation as they grow, it is expected to be randomly 
selected in the isotropic case \cite{brener-96-8}.

The morphology diagram of Brener \ea \cite{brener-96-8} 
describes the expected growth  
as surface tension 
anisotropy and undercooling are changed.  Although it is assumed to be  
a general phase diagram, 
the boundaries separating different types of growth are determined
for the solidification of a pure material, so no explicit dependence on 
solute concentration is known for doublon growth in binary alloys. 
Solute concentration does affect the undercooling \cite{reviews-PRE-01} and
must be included for a proper analysis of solutal doublons. 
Solutal undercooling is proportional to sample concentration
($= m C_\infty (1-k)/k$, where 
$m$ is the liquidus slope, 
$C_\infty$ is bulk sample concentration, and $k$ is the partition 
coefficient).
We find below that for a given grain and growth speed, 
doublons are preferred at low concentrations (small  
solutal undercooling) rather than at large undercooling as expected
for thermal doublons.

Stable ``parity broken cells'' were first observed numerically by Brener \ea 
\cite{Brener.ea:93:Crystal} and Ihle and M$\ddot{\mathrm{u}}$ller-Krumbhaar
\cite{ihle-mk-93-4}.
Since the early observation of an array of ``doublets'' by Jamgotchian \ea
\cite{Jamgotchian.ea:93:Array}, they have attracted significant interest.  
The existence of doublons was noted in eutectic growth by Kassner \ea
\cite{Kassner.ea:93:New}.
Subsequent
simulations \cite{Kupferman.ea:95:Coexistence,Kopczynski.ea:97:Cellular},
theories \cite{Brener.ea:96:Cellular,ben-amar-brener} and experiments 
\cite{Losert.ea:98:Selection,Georgelin.ea:97:Oscillatory,Ludwig:99:Dendritic} 
have probed their 
stability, characteristics and formation.  
In particular, 
Losert \ea~\cite{Losert.ea:98:Selection} 
imposed periodic perturbation experimentally to test doublon
stability.  
In their work, they didn't find doublons appearing without an imposed perturbation 
except as transients. 

In this article, we examine the doublon morphology experimentally.
In particular, we study the effects of concentration changes 
and sample orientation on the stability of 
solutal dendrites.
By studying these effects in individual grains, we are able to identify the
effects due specifically to either 
concentration or orientation.
We also study the expected transition from fractal seaweed to doublon growth 
with increasing growth rate.  We find that 
doublon selection depends strongly on concentration, with doublons 
selected at low solute concentrations (small undercooling) 
and seaweeds at higher concentrations (large undercooling). 
We also find that doublons in directional solidification 
have a particular orientation and that 
sample orientation affects the existence 
and character of the doublons.

\section{Experimental Techniques}

The experimental apparatus used presently 
has been described previously \cite{Utter.ea:02:Dynamics}
and additional details will be presented elsewhere\cite{Utter:04:Apparatus}.
We perform experiments with a  traditional directional 
solidification apparatus 
in which a quasi-two-dimensional sample 
$(13~cm \times 1.5~cm \times (5-60)~{\mu}m)$ is pulled 
through a linear temperature gradient at a constant pulling velocity.
After an initial transient, the average speed of the solidification 
front is equal to the 
pulling speed, set by
a linear stepping motor with 4 nm step size. 

The cell consists of two glass plates glued together and filled 
with the sample.  The glass plates are cleaned in stages using detergent, 
acetone, methanol, an acid solution (sulfuric acid and NoChromix), 
and distilled water.  The glue used is the epoxy Torr-Seal. 
The nominal cell depth is set by a Mylar 
spacer.

In each set of runs, we maintain the temperature gradient $G$ at 
a fixed value between 3 and 50 K/cm with a 
stability of $\pm 2$ mK.   The temperatures of the 
hot and cold sides are above and below the equilibrium melting 
temperature of $\approx 58^\circ C$
so that the solid-liquid interface remains within the gap between the 
temperature controlled blocks.
It is also possible to rotate the cell within 
the sample plane between runs.  This allows for control
over in-plane sample orientation.

The sample used is a model alloy of succinonitrile (SCN) and a small amount of 
added solute.  The solutes used in this study are
either 0.25\% poly(ethylene oxide) (PEO)\cite{peo-info},  
1.5\% acetone (ACE), or 1.3\% camphor (CAM).  
The diffusivities $D$ and partition coefficients $k$ are listed in 
Table~\ref{alloy-properties} with the solute concentrations $C$ and 
sample thicknesses $d$.

\begin{table}[tb]
\begin{ruledtabular}
\begin{tabular}{|l|c|c|c|} 
			& ACE-SCN	& CAM-SCN	& PEO-SCN  \\ 
$D$ ($\mu$m$^2$/s)	& 1270$^a$	& 300$^b$	& 80	\\
$k$			& 0.1$^a$	& 0.33$^c$	& 0.01	\\
$C$ (weight \%)		& 1.5\%		& 1.3\%		& 0.25\% \\
$d$ ($\mu$m)		& 20		& 22		& 60 \\
\end{tabular}
\caption{
Properties of samples used in this study.  
Succinonitrile alloys with acetone, camphor and poly(ethylene oxide)
as solutes.  Diffusivity $D$, partition coefficient $k$,  
solute concentration $C (\pm 0.02\%)$, and sample thickness $d (\pm 2 \mu$m)
also listed. 
$a$) from \protect \cite{Chopra.ea:88:Measurement}.
$b$) from \protect \cite{Sato.ea:87:Experiments}. 
$c$) from \protect \cite{Taenaka.ea:89:Equilibrium}.
}
\label{alloy-properties}
\end{ruledtabular}
\end{table}

We purify succinonitrile using a vacuum distillation apparatus.
Mixing with solute is performed under inert
argon atmosphere.
Samples are filled using a vacuum filling technique in order to de-gas the
sample and prevent contamination \cite{Utter:04:Apparatus}.

We observe 
the liquid-solid interface with phase contrast 
or Hoffman modulation contrast microscopy.  Sequences 
of images are recorded using a CCD camera with a framegrabber or time 
lapse video.  

To initiate growth, we melt the sample completely and 
quench it, seeding a number of grains.  
We select one grain with the desired orientation and all others
are melted off so
the chosen grain can grow and fill the width of the cell.  
To have a nearly isotropic effective surface tension within the 
growth plane, the chosen grain must be near the \{111\} plane
\cite{Akamatsu.ea:95:Symmetry-broken}.
Before each run, the sample is kept stationary ($V = 0$) 
for a sufficient time to equilibrate.


\section{Results}

Our early observations of doublons confirmed the fact that they are 
generally unstable to tilting \cite{Akamatsu.ea:95:Symmetry-broken}.  
Doublons that appear to be growing straight will eventually begin 
leaning towards one side and the cell on that side will suddenly be  
convected away.  The remaining cell then splits to form another transient 
doublon.  The tilting instability generally repeats on the same side
since a small amount of crystalline anisotropy usually breaks the 
symmetry. 
In one of our first efforts to study doublon formation, we examined 
the degenerate seaweed described previously 
\cite{Utter.ea:01:Alternating,Utter.ea:02:Dynamics}. 
At higher growth velocities, we do find the 
expected doublonic structures, 
but they are often intermittent and short-lived.  
As a doublonic dendrite forms,
there are initially two asymmetric cells with a 
characteristic gap of well defined thickness as in 
Fig.~\ref{doublon-seaweed}B.  Although there is not a parabolic envelope
over the tips, they are clearly asymmetric with the tips closer together
than the cellular spacing and the gap appears to be well selected.


\begin{figure}
\includegraphics[width=2.5in]{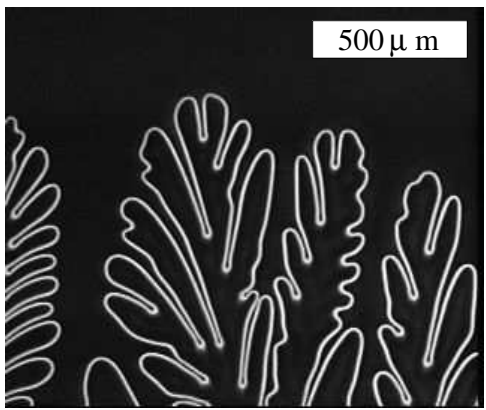}
\caption{
Degenerate seaweed at low growth speed. 
The initial tip splitting appears similar to the process of 
doublon formation. 
The sample is 0.25\% PEO-SCN at V = 2.71 $\mu$m/s and $G$ = 18K/cm.
}
\label{deg-seaweed}
\end{figure}

\begin{figure}
\includegraphics[width=3.3in]{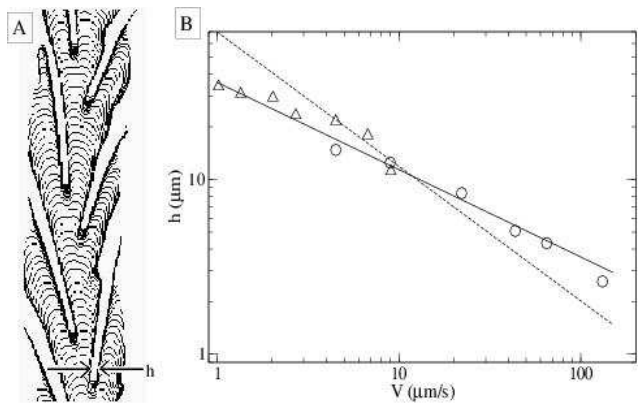}
\caption{
The evolution of the tip region over time is shown on the left.
The regularity of the gaps suggests doublon formation.  The 
gap thickness $h$ is plotted versus pulling speed for ($\triangle$)
PEO-SCN and ($\bigcirc$) ACE-SCN.  The solid line shows  
$h \propto V^{-0.5}$.  Fitting the exponent gives ($\triangle$) -0.45
$\pm$ 0.05 and ($\bigcirc$)
$-0.52 \pm 0.05$.   This compares to Brener \ea's predicted exponent 
for doublons
($h \propto V^{-7/9}$, dashed line).
}
\label{gapvsv}
\end{figure}

Doublons are predicted to be the basis for seaweed growth, so it is
no surprise that the same description of the tilt instability 
could be given of seaweed tip splitting.
During the tip splitting of seaweeds, the cell briefly appears as a pair 
of asymmetric cells as in Fig.~\ref{deg-seaweed}.  This is the same 
grain as in Fig.~\ref{doublon-seaweed}B at a lower growth speed.
The gap $h$ between 
the fingers is also quite regular.  To show this more clearly,
in Fig.~\ref{gapvsv}A, the interface near the tip region 
has been extracted in subsequent images 
and displaced upwards a distance $V \Delta t$ where $\Delta t$ 
is the time interval
between pictures, i.e. this is an image in a frame where the tip grows 
upwards at speed $V$.  There we see that the initial gap between seaweed lobes
is clearly selected, as with doublons.  In Fig.~\ref{gapvsv}B, 
the gap thickness is
measured for seaweeds at different pulling speeds.  
The gap thickness scales approximately as $h \propto V^{-0.5}$.  
The scaling for the seaweed gap thickness 
is consistent with the $\lambda \propto V^{-0.5}$ scaling found for 
fingering wavelengths in solidification \cite{Trivedi.ea:94:Solidification}
and inconsistent with the 
prediction of $h \propto V^{-7/9}$ for 
doublons \cite{brener-96-8}.  
The discrepancy could indicate that the seaweed tip is not strictly 
a transient doublon and that doublons are not the fundamental 
building blocks of seaweed growth.  However, we are not able to verify 
the 7/9 exponent for the solutal doublons we observe, since 
the resolution of our images is not sufficient 
to resolve the doublon gap thickness.

\begin{figure}
\includegraphics[width=3.3in]{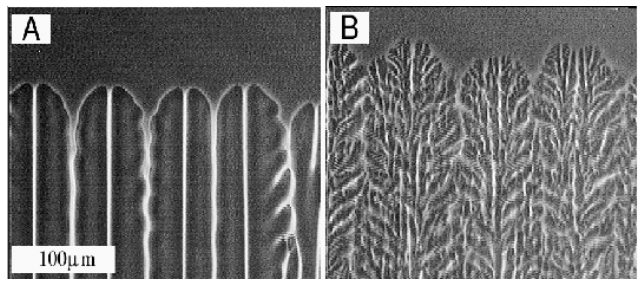}
\caption{
Doublon to seaweed transition with an increase in concentration. 
The same grain is grown from a (A) zone refined (low concentration) into a 
(B) bulk (high concentration) region of the same cell.
The sample 
is CAM-SCN at V = 86.4 $\mu$m/s and $G$ = 40 K/cm.
}
\label{conctrans-CAM}
\end{figure}

\begin{figure}
\includegraphics[width=3.0in]{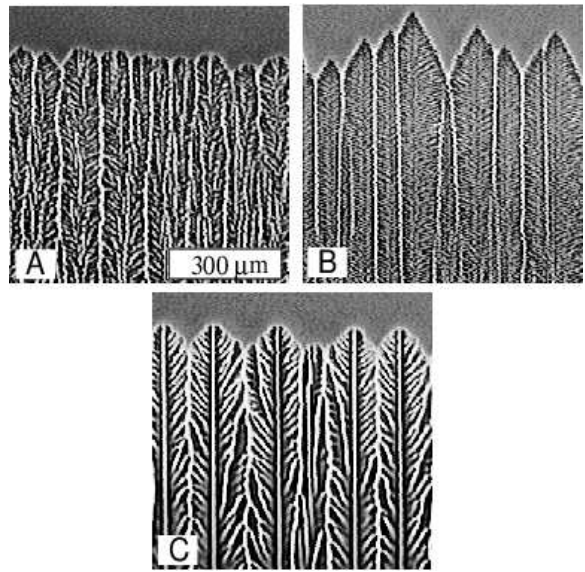}
\caption{
(A) Seaweed to (C) doublon transition with a decrease in concentration. 
During this transition, the interface rapidly advances, forming 
(B) transient superdendrites before forming doublons.  The sample 
is CAM-SCN at V $\approx 150 \mu$m/s and $G$ = 18 K/cm.
}
\label{conc-transition-SD}
\end{figure}

We find the stability of doublons to be strongly concentration dependent.
In a seaweed grain showing unstable doublons, we allowed a 
flat interface to grow at small pulling speeds, effectively zone-refining
a section of the cell.  After backing up and growing through the 
zone refined area and into a region of higher concentration, we observe
stable doublons which break apart into unstable seaweed structures as 
the solute concentration increases, 
as in Fig.~\ref{conctrans-CAM}.  This is repeatable for the different 
mixtures used in this study.  
In one case, we then rotated the cell 
$180^\circ$ and forced the grain to grow from a region of high concentration
to low concentration and saw the opposite transition from unstable 
seaweeds to stable doublons.   
The latter transition
is shown in Fig.~\ref{conc-transition-SD}.
Note that since the average interface position is at a higher melting 
temperature for lower solute concentration, the camera is moved along 
the growth direction 
and the relative thermal 
undercooling cannot be determined directly 
from these images. 

\begin{figure}
\includegraphics[width=3.3in]{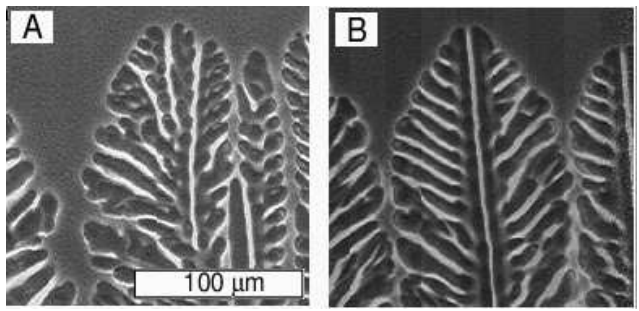}
\caption{
Seaweed to doublon transition with a decrease in concentration. The sample 
is CAM-SCN at V $\approx 150 \mu$m/s, and $G$ = 18 K/cm.
}
\label{conc-transition}
\end{figure}

In the seaweed to doublon transition, since the interface must advance 
quickly towards a new equilibrium interface location as the 
concentration decreases, we see transient superdendrites, shown in 
Fig.~\ref{conc-transition-SD}B. 
These are triangular growths that commonly form at large growth 
velocities \cite{Ragnarsson:1998:Super}.  The progression is counter 
to what might be expected from Brener \ea's morphology diagram.  At lower
concentration, the interface advances to lower undercooling, so we 
might expect a transition from doublon to seaweed growth, contrary to 
our observations.  However, as mentioned earlier, the morphology diagram 
is discussed in the context of a pure sample and may not be applicable 
with concentration changes. 
We show a second example of the seaweed to doublon transition in 
Fig.~\ref{conc-transition}.

It is interesting to note that the unstable seaweed growth in  
each case still maintains a coarse spacing that is similar to that for the 
doublons.  The doublons we observe though are not alternating shallow
and deep grooves as in \cite{Losert.ea:98:Selection}.  Particularly in 
Fig.~\ref{conc-transition}B, we see dendritic doublons with much more 
strongly 
developed sidebranches
than most previous observations 
\cite{Akamatsu.ea:95:Symmetry-broken}.

At low concentrations, doublons appear to be stable and 
rarely undergo the tilting instability described above. 
Doublons appear to be strongly selected without an imposed 
modulation (e.g. Fig.~\ref{conctrans-CAM}) unlike  
what is reported by Losert \ea \cite{Losert.ea:98:Selection}.

In numerical 
simulations, higher noise deters doublon growth
\cite{Losert.ea:98:Selection}.  The relevant noise in our system 
is most likely concentration fluctuations rather than the thermal noise 
relevant in growth of pure materials.  In this case, higher concentrations 
would then correspond to larger fluctuations and larger noise.
It is also possible that the surface tension changes with concentration
\cite{Charach.ea:99:Phase}
which could lead to a change in morphology.
This seems plausible because higher concentrations
(and thinner samples) lead to seaweed growth more readily than dendritic
growth, particularly when using PEO as solute.

\begin{figure}
\includegraphics[width=2.9in]{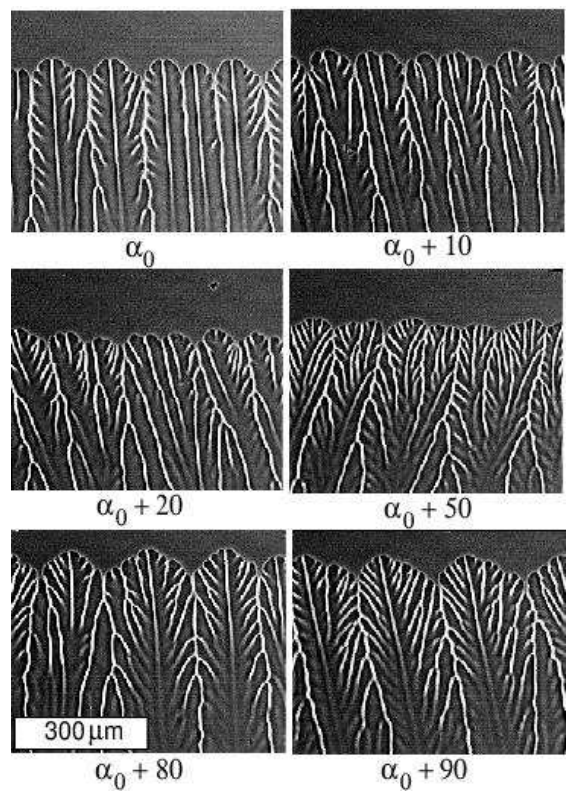}
\caption{
Doublon growth with changes in orientation.  A grain showing doublon growth
shows transient doublons when rotated counterclockwise
by $10^\circ$ and essentially no
doublon formation at higher rotation angles.  The sample is 0.5\% PEO-SCN
at V = 86.4 $\mu$m/s.
}
\label{doub-direc-all6}
\end{figure}

We also observe an orientational dependence of doublon growth.  It is 
believed that for completely isotropic systems, doublons will 
spontaneously select a growth direction since none is preferred 
\cite{brener-96-8}.  However, Losert \ea~report that no stable doublons
are found for zero anisotropy in simulations \cite{Losert.ea:98:Selection}.  
In our experiments,
anisotropies are clearly present and lead to a preferred doublon orientation.
Fig.~\ref{doub-direc-all6} 
shows the results of an experiment 
in which the cell can be rotated within the
sample plane in order to 
further probe the effect of crystal orientation on growth.  
The arbitrary angle $\alpha_0$ reflects the 
fact that we know relative angles as the sample is rotated rather than 
absolute crystallographic orientations.  We see 
that doublons are oriented along particular directions 
and their stability depends on orientation.
In particular, doublons are more stable when oriented along a favored growth 
direction.  One explanation for this is that the tilt instability 
described above is more prevalent for tilted doublons.  We also note 
that the crystalline anisotropy need not be fourfold symmetric as 
is often assumed \cite{Utter.ea:02:Dynamics}.


\begin{figure}
\includegraphics[width=1.5in]{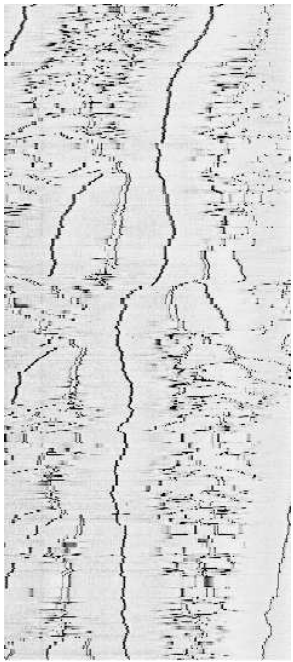}
\caption{
Space-time plot of doublon growth in 0.5\% PEO-SCN at V = 22 $\mu$m/s.
This run corresponds to the sample shown in Fig.\ref{doublon-seaweed}.
The dark central line indicates the inner groove of a doublon
as it meanders in the field of view.
The image width is 243 microns and the time of 48 seconds increases upwards. 
}
\label{doub-ST}
\end{figure}

The long-time behavior of doublon growth can be seen in Fig.~\ref{doub-ST},
which shows a space-time plot for doublons shown in 
Fig.~\ref{doublon-seaweed}A.  
This is essentially a chart recording of 
the growth and is created by extracting lines 
at a fixed distance behind the interface for sequential pictures
(see~\cite{Akamatsu.ea:95:Symmetry-broken}, for example).  
The central gap of the doublon is identified by the dark line.
Although we observe a doublon tip 
growing for over 48 seconds (in a run of several minutes), 
it meanders over time.  This is different from the oriented 
growth of dendrites which typically grow along fixed, preferred directions
at steady-state. Doublons in this regime are also 
eliminated through the tilt instability, but their lifetime is significantly
longer.  The gap between fingers is observed to be very uniform over time.  
The positions of the tips of the asymmetric pair 
along the growth direction is also equal within 
the resolution of our data. Therefore, we are not able to test 
M$\ddot{\mathrm{u}}$ller-Krumbhaar \ea's
suggestion for the mechanism of doublon stability 
\cite{Muller-Krumbhaar.ea:96:Morphology}
in which if one finger grows ahead, 
it has more room and widens, leading to a lower growth velocity.

\begin{figure}[tb]
\includegraphics[width=2.5in]{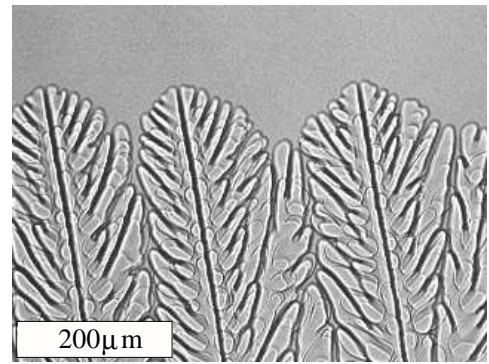}
\caption{
Doublonic dendrites.  Sidebranches perpendicular to the plane of growth 
are evident.  The sample is PEO-SCN at V = 86.4 $\mu$m/s.
}
\label{sidebranching}
\end{figure}

Fig.~\ref{sidebranching} shows doublonic dendrites with sidebranches 
perpendicular to the sample plane.
In this case, the 
sample thickness is large enough to support the transverse 
sidebranching mode. 
In three dimensions, the triplon is predicted to be the basic building block
of isotropic growth \cite{Abel.ea:97:Three-dimensional}, although transient
doublons have been observed in 3D growth of 
xenon dendrites \cite{Stalder.ea:01:Morphology}.  If we
were able to increase the thickness of the sample, we would expect a 
transition from the three-dimensional doublons in Fig.~\ref{sidebranching}
to triplons, although the nature of the transition is unknown. 
In two dimensions, multiblons have been 
shown in numerical simulations and were argued to be generically 
unstable \cite{Kopczynski.ea:97:Cellular}.
In our observations, multiblons can be seen as transients but are
not found to be stable.

\section{Conclusions}

In conclusion, we observe doublons in low anisotropy 
growth but they are 
often unstable to a tilting instability. 
We find doublon formation to be strongly dependent
on solute concentration and sample orientation. 
Doublons are selected at low concentrations (small solutal undercooling)
in contrast to the fact that doublons exist at large 
thermal undercoolings in 
pure materials.  Perhaps the dominant factor is that larger concentrations
lead to larger fluctuations which destabilize the tip.
Doublons also exhibit a preferred growth direction and changing the 
orientation of a specific doublonic grain demonstrates that the 
character and stability of doublons depend on crystalline orientation. 
Even when stable, doublons tend to meander, unlike in dendrite growth.
At higher concentrations and when 
the preferred growth direction is substantially
different from the imposed growth direction, 
we observe the seaweed morphology.  
We observe seaweed-doublon transitions with changes in these parameters.
In seaweed growth, the tip splitting process appears similar to the formation 
of transient doublons.  However, the gap thickness $h$ scales with the 
velocity $V$ as $h \propto V^{-0.5 \pm 0.05}$ rather than $V^{-7/9}$ as 
predicted for doublons.  This might indicate that seaweed growth should 
not strictly be viewed as composed of transient doublons.
However, with the resolution of our images, we are not able to verify or 
reject the 7/9 exponent for the narrow gap thickness of solutal doublons. 

It remains unclear what determines the  
stability of solutal doublons, 
particularly with changes in solute concentration.
Doublons should also be observable in other systems if the assumed 
morphology diagram  
is generic.  The effects of anisotropies or concentration 
in these systems are open questions.

{\bf Acknowledgement} 
We would like to thank Rolf Ragnarsson for significant contributions 
in related work leading to these experiments.
This work was supported by the Cornell Center for Materials Research (CCMR), 
a Materials Research Science and Engineering Center of the National Science
Foundation (DMR-0079992).



%
%

%
%

\end{document}